\documentclass[aps,pre,a4paper,twocolumn,showpacs]{revtex4}
 
\usepackage{graphicx,amsmath,amssymb,latexsym}
\begin{document}

\title{Effects of small surface tension in Hele-Shaw multifinger 
dynamics:\\
an analytical and numerical  study}

\author{E. Paun\'e$^1$, M. Siegel$^2$, and J. Casademunt$^1$
}

\affiliation{
$^1$Departament d'Estructura  i Constituents de la Mat\`eria,
Universitat de Barcelona,\\
Avinguda Diagonal 647, 08028 Barcelona, Spain\\
$^2$Mathematics Department, New Jersey Institute of Technology, Newark,
NJ 07102, USA
}

\begin{abstract}
We study the singular effects of vanishingly small 
surface tension on the dynamics of finger competition
in the Saffman-Taylor problem, using the asymptotic techniques
described in [S. Tanveer, Phil. Trans. R. Soc. Lond. A {\bf 343}, 155 (1993)]
and [M. Siegel, and S. Tanveer,
Phys. Rev. Lett. \bf76\rm, 419 (1996)]
as well as  direct numerical 
computation, following the numerical scheme of 
[T. Hou, J. Lowengrub, and M. Shelley,
J. Comp. Phys. \bf 114 \rm, 312 (1994)].
We demonstrate the dramatic effects of small surface tension on the late time
evolution of two-finger configurations with respect to exact (non-singular) 
zero-surface 
tension solutions.  The effect is present even when
the relevant  zero surface tension solution
has asymptotic behavior  consistent with selection theory. 
Such singular effects therefore cannot be traced back to 
steady state selection theory, and imply a drastic 
global change in the structure of phase-space flow. 
They can be interpreted in the framework of a 
recently introduced dynamical solvability scenario according to which 
surface tension unfolds the structually unstable flow, restoring
the hyperbolicity of multifinger fixed points.

\end{abstract}

\maketitle

\section{Introduction}

The displacement of a viscous fluid by a less-viscous one in a Hele-Shaw
cell, the so-called Saffman-Taylor problem 
\cite{Saffman58,Bensimon86a,McCloud95,Tanveer00,Casademunt00},
is a prototypical pattern formation problem.
Since the
seminal work of Saffman and Taylor \cite{Saffman58}
a considerable effort has been aimed at understanding
both steady and unsteady interfacial patterns 
formed during this flow.
The Saffman-Taylor problem
is the simplest member of a wide class of interfacial pattern formation
problems such as free dendritic growth,
directional solidification,
or chemical electro-deposition \cite{Langer87,Pelce88,Kessler88}.
As such, a theoretical understanding of Hele-Shaw flow may
help elucidate generic behavior common to many pattern forming systems. 
Despite its relatively
simple formulation and the large amount of work devoted to it, 
however, several
aspects of interfacial dynamics in Hele-Shaw flow
are still poorly understood, in particular 
concerning the highly nonlinear and nonlocal dynamics of finger competition.
 
One of the main reasons  for the wide interest
in Hele-Shaw flow, at least  from a mathematical
point of view, is that explicit time-dependent solutions can be found 
in the case of 
zero surface tension
\cite{Howison86,Dawson94,Entov95,Magdaleno00}.
However, it is also known that 
the zero surface tension Saffman-Taylor (ST) problem is ill-posed
as an initial value problem \cite{Tanveer93} and finite-time singularities
appear frequently \cite{Howison85}. Nevertheless, rather large classes 
of zero surface tension
solutions have been found which exhibit the variety of morphologies
observed both in experiments and numerical simulations.
Then, the question that naturally arises is to what extent smooth 
(nonsingular) 
zero surface tension solutions reproduce the dynamics of the
physical problem with finite surface tension, in particular
in the limit of vanishing dimensionless surface tension, $B \rightarrow 0$.
 
It is well known that surface tension is a singular perturbation
to the zero surface tension problem \cite{Tanveer93}.
This singular character 
manifests dramatically in the classical selection problem 
posed by Saffman and Taylor \cite{Saffman58} 
and only solved three decades later
\cite{Shraiman86,Hong86,Combescot86,Tanveer87},
where an arbitrarily small surface tension selects out
a single, stable solution from the continuum of steady single-finger $B=0$
solutions. Another manifestation of the
singular nature of surface tension which is directly relevant to the 
present work is its effect on the dynamics.
Siegel, Tanveer and Dai \cite{Siegel96a,Siegel96b}
showed that interfacial evolution for  the regularized problem
(i.e., vanishingly small $B$) may differ signifficantly from
that for the $B=0$ problem in order one time. Then, smooth time dependent
solutions of the $B=0$ case do not coincide, in general, with
the limiting solutions for  $B \rightarrow 0$. Accordingly, 
the study of finite $B$ dynamics
encounters considerable difficulties since $B=0$ solutions cannot
be naively  used as a starting point for the study of
the problem with finite $B$.

The physical content of exact zero surface tension solutions with pole-like
singularities has been recently addressed in 
Refs.~\cite{Casademunt00,Magdaleno98,Paune02a}
using a dynamical systems approach. 
Through a detailed study  
it has been shown that the exact zero-surface tension phase flow, 
considered in a global sense, is 
structurally unstable. 
In other words, the zero surface tension phase dynamics  are 
\it not \rm topologically equivalent to the phase space flow of the 
physical problem, regularized by surface tension. Indeed, 
the zero surface tension phase flow omits the 
necessary saddle-point structure of multifinger fixed points, which 
is crucial 
to the physical finger competition process~\cite{Paune02a}.
A natural extension of the well known solvability mechanism (first applied
to  `select'
a finger of width 1/2 out of a continuum of solutions in
the single finger case)
was proposed  for multi-finger solutions in~\cite{Paune02a}; 
this helps clarify how the introduction
of surface tension modifies the global phase space structure 
of the flow and restores the hyperbolicity of
multi-finger fixed points.

The approach of Ref.~\cite{Paune02a}, however, was qualitative in nature and 
could not quantify the extent to which zero surface tension  trajectories  
might resemble the evolution with small surface 
tension. In particular it was recognized that, while some trajectories 
appear to be qualitatively correct for infinite time, others may have 
a dramatically different evolution. In particular, adding an infinitessimal
surface tension could give the opposite outcome in the  finger competition, 
that is, make the `losing' finger for $B=0$  become
the `winning' finger when $B>0$,  for sufficiently generic 
sets of initial conditions.

A satisfactory analytical understanding of the problem with $B \neq 0$
has been achieved in two regimes: the initial linear instability
of the flat interface followed by the weakly non-linear regime 
\cite{Alvarez01},
and the asymptotic regime where surface tension selects the width
of the single finger \cite{Shraiman86,Hong86,Combescot86,Tanveer87}.
The highly non-linear intermediate regime
that connects the quasi-planar interface with the asymptotic
single-finger regime has mostly been studied through 
numerical computation. The first exhaustive numerical studies were 
reported by
Tryggvason and Aref \cite{Tryggvason83,Tryggvason85}, who
paid considerable attention to the influence of viscosity
contrast on the problem and studied both single-finger
and multi-finger configurations. Later, Casademunt and Jasnow 
\cite{Casademunt91,Casademunt94}
showed that the basin of attraction of the single-finger solution 
depends strongly on viscosity contrast and that only when one of the two 
fluid viscosities is negligible it can be claimed that 
the single finger is the 
universal attractor of the problem. In the present work we will restrict 
to this limiting case. 
DeGregoria and Schwartz
\cite{DeGregoria85,DeGregoria86} observed that well-developed
fingers split when surface tension is sufficiently decreased.
This tip-splitting instability is related to the fact that
the Saffman-Taylor finger is linearly stable but non-linearly
unstable, and the size of the perturbation that triggers the
tip-splitting decreases quickly with surface tension
\cite{Bensimon86b}.
Dai and Shelley~\cite{Dai94} showed that for small $B$
numerical computations are extremely sensitive to the
precision used in the computations. As a consequence
noise level has to be controlled with care in order to
ensure that the computation is sufficiently accurate.
Hou \it et al. \rm \cite{Hou94} developed a numerical method that deals with
the numerical stiffness of the problem in an efficient manner, aiding
the ability to perform long time computations.
More recently Ceniceros \it et al.\rm~\cite{Ceniceros99,Ceniceros00},
using very high precision arithmetic have been able to study
the effect of extremely small surface tension in the circular
geometry with suction, and they have observed that surface
tension can produce complex ramified patters even without the
presence of noise.
 
An  analytical treatment of  this highly nonlinear and nonlocal 
free-boundary problem   
faces challenging 
difficulties. In particular, a perturbative  study
for small $B$   
is complicated by the ill-posedness of the zero
surface tension  problem. Tanveer \cite{Tanveer93}
was able to overcome this obstacle 
by embedding the zero surface tension problem in a well-posed
one. In addition, this well-posed extension of the $B=0$ problem
allowed Baker \it et al.\rm~\cite{Baker95} to develop a numerical
method to compute the time evolution of zero surface tension dynamics 
in a well-posed manner.
Once the $B=0$ problem is formulated in a well-posed way 
the $B \neq 0$ case can be studied using a perturbative approach.
The main result of the asymptotic perturbative theory developed
by Tanveer \cite{Tanveer93} is that the effect of surface tension
may be manifest in a $O(1)$ time: the evolution of
the same initial interface for $B=0$ and
$B \neq 0$ will in general differ after a time
of order one, even if the $B=0$ solution is smooth for all time.
Siegel \it et al.\rm~\cite{Siegel96b} have extended
the work of Ref.~\cite{Tanveer93} to later stages of the evolution,
and through numerical computation for very small values of $B$
they showed that smooth $B=0$ solutions are indeed significantly
affected by the presence of arbitrarily small $B$ in order-one time,
thus confirming the predictions of the perturbative theory.
The zero surface tension solutions studied by Siegel 
\it et al.\rm~\cite{Siegel96a,Siegel96b} in the channel 
geometry were single-finger solutions
with an asymptotic width $\lambda$, 
specifically chosen to be  incompatible with selection
theory for vanishing surface tension. They found that the singular
effect of surface tension was to widen the finger in order to
reach the selected width.
The surprising feature  here is that the effect of surface tension is
felt in order-one time, i.e., that the time lapse for which the 
regularized solution approaches the unperturbed one as
$B \rightarrow 0$  is bounded.
 
The present paper expands the work of Refs.~\cite{Siegel96a,Siegel96b}
in the spirit of Ref.~\cite{Paune02a}, 
towards the study of multi-finger solutions. However,
unlike the studies of \cite{Siegel96a,Siegel96b}
we chose zero surface tension multi-finger soltuions  which are compatible with
asymptotic selection theory, that is, with an
asymptotic finger width $\lambda=1/2$---the selected value
in the limit $B \rightarrow 0$. 
In this way we isolate the intrinsic finger competition dynamics from the 
selection effects responsible of restoring the asymptotic width.
Two different kinds
of two-finger zero surface tension solutions are studied,
and in both cases it is shown that surface tension acts
as a singular perturbation to the dynamics in order-one time,
modifying dramatically the late time configuration of
the interface not only quantitatively but also qualitatively.
Specifically we show that paths in phase space associated with
zero and nonzero surface tension evolution, 
and indeed the global topological structure of the
the phase spaces, may differ appreciably, even for arbitrarily
small $B$.   
In physical terms, our evidence suggests 
that the 
presence of arbitrarily small
surface tension can completely alter outcome of finger competition
when compared with the zero surface tension evolution. 
 
The paper is organized as follows:
in Sec.~\ref{zero_b} the equations describing Hele-Shaw flow are
introduced, and a class of two-finger zero surface tension solutions
relevant to two-finger competition is presented and briefly
discussed. This class of solutions will be used as initial condition
for numerical computation with $B > 0$. In Sec.~\ref{Asymp}
the basic features of the asymptotic theory are recalled,
and  the theory is applied to the zero surface tension
solutions introduced in the previous section. The numerical computations
with finite (but small) $B$ are presented in Sec.~\ref{numerics}.
Sec.~\ref{concl} discusses and summarizes the results obtained in
previous sections.

\section{Zero surface tension}
\label{zero_b}

In this section we present the equations which govern the interfacial
dynamics in a rectilinear Hele-Shaw cell, following the formalism 
of  \cite{Tanveer93}. We consider a class of  exact, time-dependent
 zero surface tension 
solutions that are relevant to the finger competition
problem, and briefly describe the solutions within this class.

Consider  Hele-Shaw flow in the channel geometry, 
in which a fluid of negligible
viscosity displaces a viscous liquid.
The equations governing the interfacial evolution can be 
conveniently formulated by
first  introducing  a conformal map $z(\zeta,t)$ 
which takes the interior of the unit semicircle
in the $\zeta$ plane into the region occupied by the viscous fluid
 in the complex plane $z=x+{\rm i}y$, 
in such a way that the arc $\zeta=e^{{\rm i}s}$
for $s \in [0,\pi]$ is mapped to the interface 
and the diameter of the semi-circle
is mapped to the channel walls\footnote{For interfaces symmetric with respect
to the central axis of the channel our formulation also 
describes periodic boundary
conditions, that is, an infinite array of fingers.}. 
The mapping function $z(\zeta,t)$ has the form 
$z(\zeta,t)=-(2/\pi) \ln \zeta +{\rm i} +f(\zeta,t)$, 
and inside and on  the unit semi-circle we require 
$f(\zeta,t)$ to be analytic and $z_{\zeta}(\zeta,t) \neq 0$.
In addition, we require that
\begin{equation} \label{symm}
{\rm Im}~{f}=0 
\end{equation}
on the real diameter of
the semi-circle.  This latter condition  ensures that 
$z$ maps the diameter to the channel walls.
Under suitable assumptions (see \cite{Tanveer93}) the 
Schwartz reflection principle may
be applied to show  that $f$ is analytic
and $z_\zeta \ne 0$ for $|\zeta| \leq 1$.  

The effective velocity field,  averaged across the plate gap, is a 
two-dimensional potential flow satisfying Darcy's law
\begin{eqnarray}
{\bf u}=\nabla \varphi.
\end{eqnarray}  
Here  $\varphi$ is a velocity potential defined by  
$\varphi=-(b^2/12\mu)~p$, where $p$ is the pressure, $\mu$ is the viscosity
and $b$ is the gap width.  Under the assumption of  
incompressibility $(\nabla\cdot{\bf u}=0)$
the potential satisfies  Laplace's equation
\begin{eqnarray}
\nabla^2 \varphi=0.
\end{eqnarray}
Incompressibility also  implies the existence of a stream function 
$\psi$. Therefore,
one can define
a complex velocity potential $W(z,t)=\varphi + {\rm i}\psi$ which
is analytic for $z$ in the fluid region of the channel.
Its form as a function of $\zeta$ reads
\begin{eqnarray}
W(\zeta,t)=-(2/\pi)\ln \zeta + {\rm i} + \omega(\zeta,t) 
\end{eqnarray}
where  $\omega(\zeta,t)$ is an analytic function inside the unit circle. 
The condition that there is no flow through the walls implies that 
${\rm Im}~{\omega}=0$ must hold on the real diameter of the  unit semi-circle.
In the absence of surface tension, $\omega=0$ (see Eq.~(\ref{evol2})).

At the interface we impose the usual boundary conditions. 
The kinematic boundary
condition states that the normal component of fluid
velocity at a point  on the interface equals the normal velocity
of the interface at that point, and takes the form
\begin{equation} \label{evol1}
{\rm Re}\left[ \frac{z_t}{\zeta z_{\zeta}}\right]=\frac{1}{|z_{\zeta}|^2}
{\rm Re}\left[\zeta W_{\zeta}\right].
\end{equation}
The dynamic boundary condition specifies that the pressure jump
across the interface is balanced by surface tension, and
is given by
\begin{equation} \label{evol2}
{\rm Re}~ \omega=
-\frac{B}{|z_{\zeta}|}{\rm Re}\left[1+\zeta \frac{z_{\zeta\zeta}}
{z_{\zeta}}\right].
\end{equation}
The parameter $B$ is the nondimensional surface tension and is defined by
$B=b^2T/12\mu Va^2$, where $T$ is
the surface tension, $V$ is the fluid velocity at infinity and $a$ is half the 
cell width.  The equations given in (1)-(5) are in nondimensional form, 
with lengths and velocities nondimensionalized by  $a$ and $V$, respectively.

When $B=0$ it is well known 
that pole singularities in $z_\zeta$ (i.e. in $f_\zeta$)   present in 
the exterior of the unit disk  are preserved
under the dynamics, i.e., such singularities are neither 
created nor destroyed, although the 
location
 of those which are initially present will evolve with time. 
Exact $B=0$ solutions consisting of a collection of pole singularities with
constant amplitude have been the focus of extensive
studies
(see e.g. \cite{Howison86} ).
The simplest such solution leading to nontrivial finger competition
consists of  a pair of  singularites in the upper halfplane of
$|\zeta|>1$, located at positions that are symmetric with respect
to the y-axis.  A second pair of poles conjugate to the first
pair is required to satisfy the symmetry restriction~(\ref{symm}). 
This exact solution
takes the form  \cite{Casademunt00,Magdaleno98,Paune02a}
\begin{eqnarray} \label{fepsilon}
z(\zeta,t)=- \frac{2}{\pi}\ln \zeta
+ \frac{1}{\pi} (1-\lambda+i \epsilon) 
\ln (1-\frac{\zeta^2}{\zeta_s(t)^2}) \nonumber
\\
 +\frac{1}{\pi} (1-\lambda-i \epsilon) 
\ln (1-\frac{\zeta^2}{\bar{\zeta}_s(t)^2}) 
+ d(t)+ i 
\end{eqnarray}
where $\lambda$ and $\epsilon$ are real constants with $0 < \lambda < 1$
and $\epsilon \geq 0$, and $d(t)$ is real. The singularity locations are given
by  the complex parameter  $\zeta_s(t)$, which  satisfies a simple 
differential equation given in \cite{Magdaleno98}. Analyticity of $f(\zeta,t)$ 
in the unit circle implies that $|\zeta_s(t)|  > 1$.
We employ the convention
that $\zeta_s(t)$ is a complex number in the first quadrant.
The amplitudes  of the singularities, given here by the numbers
$1-\lambda+i \epsilon$ and its conjugate, are chosen  so that the 
asymptotic form of the solution
consists of one or two steadily propagating fingers of total width $\lambda$.
The parameter  $\epsilon$ determines the nature of the 
finger competition for $B=0$.

The  solution~(\ref{fepsilon}) describes generically two different
fingers of the nonviscous phase penetrating the viscous one. In the linear
regime $|\zeta_s(t)| >> 1$ the interface consists of a single bump or finger,
and as time increases a second finger may develop and grow, 
depending on the value of
${\rm Arg} \: \zeta_s(0)$.

\begin{figure}
\includegraphics*{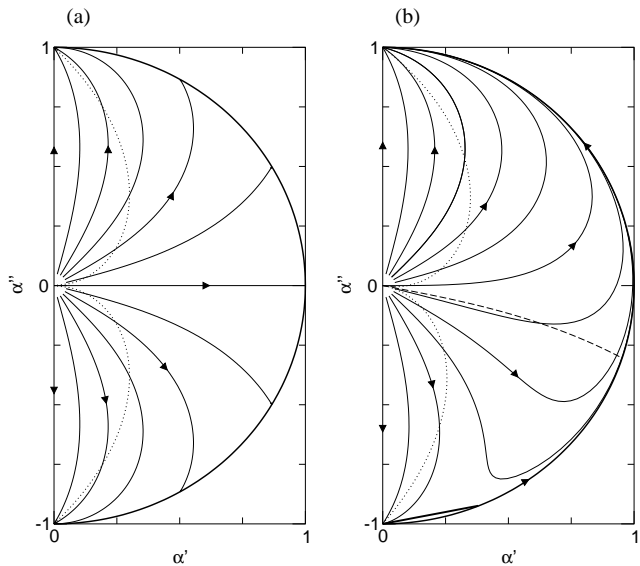}
\caption{\label{fig:phase} (a) Phase portrait for 
$\lambda=1/2$ and $\epsilon=0$. The region
between the dotted lines corresponds to two-finger configurations and
the other two regions to single-finger configurations.
(b) Phase portrait for $\lambda=1/2$ and $\epsilon=0.1$. 
For the points on the dashed
line the two fingers have the same length.}
\end{figure}

We summarize the features of the solution~(\ref{fepsilon}) 
that are most relevant to the study of finger 
competition.  Consider first   $\epsilon=0$. In this case  the
asymptotic configuration consists of one or two  fingers of total width 
 $\lambda$, depending on the initial condition.  The singularities move 
toward the unit disk, with  the limit
as $t \rightarrow \infty$ denoted  by 
$\zeta_s(t) \rightarrow e^{i \theta}$.  When $\theta=0$
the asymptotic configuration is a single Saffman-Taylor
finger growing in the center
of the channel (this asymptotic configuration is denoted ST(R)),
for $\theta=\pi/2$ it is a  `side' Saffman-Taylor finger
i.e. a pair of  half  fingers of  total width $\lambda$ with tips located
at the cell walls (denoted ST(L)), 
and for $\theta=\pi/4$ it is a `double' Saffman-Taylor
finger, namely two identical fingers of width $\lambda/2$ with tips 
at $x=0, \pm 1$ (denoted 2ST).   For any
other value of $\theta$ the 
asymptotic configuration consists  of two unequal steadily  growing fingers.  
The two-finger asymptotic configuration is a consequence of the continuum
of fixed points that is present in the phase portrait of the dynamical
variables which govern the shape of the interface, 
namely $({\rm Re} \zeta_s(t),
{\rm Im} \zeta_s(t))$.   
In order to correspond to the notation of  \cite{Magdaleno98}
 introduce the
variable $\alpha(t)=\alpha'(t)+{\rm i}\alpha''(t)=1/(i ~\zeta_s^2(t))$. Then
the planar interface corresponds to $\alpha=0$, 
the  center Saffman-Taylor finger  
to $\alpha=- i$, the side Saffman-Taylor finger to  $\alpha=i$, 
 and the double Saffman-Taylor finger to $\alpha=1$. 
Figure~\ref{fig:phase}  shows the phase portrait of the 
dynamical system obtained 
from the substitution of the mapping defined by Eq.~(\ref{fepsilon}) into the 
evolution Eqs.~(\ref{evol1},\ref{evol2}) for $B=0$, 
using  the dynamical variables 
$(\alpha',\alpha'')$ . The asymptotic states with two advancing 
fingers corresponds 
in the dynamical system $(\alpha',\alpha'')$ to a continuum of fixed points
given by $|\alpha|=1$.
Therefore, for $\epsilon=0$ the solution~(\ref{fepsilon}) does not exhibit 
finger competition.
In addition, it is important to note that  the evolution of~(\ref{fepsilon}) 
with $\epsilon=0$ is free of finite time singularities, i.e., $z_\zeta \neq 0$
in the domain  $|\zeta|  \leq 1$ for all time. \\

For $\epsilon \neq 0$ the continuum of fixed points is removed 
(see Fig.~\ref{fig:phase}),
as is the double Saffman-Taylor finger fixed point. 
Consequently, the solution to Eq.~(\ref{fepsilon}) exhibits `successful' 
competition, in the sense that the asymptotic interface shape consists of  a 
single Saffman-Taylor finger or side Saffman-Taylor finger.
The price to pay is the appearance of finite 
time singularities for a certain subset of initial conditions, in the form
of a zero of $z_\zeta$ impacting the unit disk (this is a generic feature
of conformal map solutions $z_\zeta$ composed of a finite number of pole
singularities---see~\cite{Howison86}).  Then,
only the subset of initial conditions free of finite time singularities is 
capable of sustaining finger competition all the way
to the $t \rightarrow \infty$ outcome.  Nevertheless, one may ask whether
the class of $B=0$ solutions that are free of finite time singularities may
describe, at least qualitatively, the physical finger competition
for positive surface tension in the limit $B \rightarrow 0$.

In the following sections we focus
on the class of initial data for 
which the $B=0$ solutions   are devoid of   finite
time singularities, and  examine the $B>0$ dynamics.  We develop
a general theory  for how the presence of positive surface tension
affects the outcome of finger competition. This will
enable us to predict the winner of finger competition, i.e.,  the eventual
asymptotic state.   Most interestingly, we find instances in which 
the presence of arbitrarily small surface tension 
leads to dramatically different
outcomes in finger competition when compared with the zero surface tension
evolution.

\section{Asymptotic theory}
\label{Asymp}

Little is known about the effect of finite (but small) surface tension $B$ 
on the dynamics of zero surface tension  multifinger solutions, and 
in particular on the class 
of exact solutions~(\ref{fepsilon}). 
For single finger configurations, steady state selection
theory predicts that the finger cannot have an arbitrary width. Indeed,
for vanishing surface tension $B \rightarrow 0$ the width $\lambda=1/2$ is
selected, asymptotically in time. Thus,  it is clear that surface tension
has a critical influence on  single finger solutions with $\lambda \neq 1/2$. 
The nature
of this influence in the limit $B \rightarrow 0$ has been investigated 
by  Siegel, Tanveer and Dai
\cite{Siegel96a,Siegel96b}, who present evidence that  zero surface tension 
single finger solutions with $\lambda < 1/2$ are significantly perturbed
by the inclusion of an arbitrarily small amount of surface tension in order
one time. The effect of surface tension is an increase of the finger width
to reach the width predicted by selection theory.

Consider now the effect of small surface tension on the exact ($B=0$) 
two finger solution~(\ref{fepsilon}).
When $0<B \ll 1$ the asymptotic perturbation theory developed in 
Refs.~\cite{Siegel96a,Tanveer93,Siegel96b} can be applied.
This perturbation theory describes the effects of the introduction of 
a small amount of surface tension on initial  data  $z(\zeta,0)$
specified in the extended complex plane, i.e., 
in a domain including the `unphysical'
region $|\zeta| > 1$ (the extended domain is required
to make the $B=0$ problem well-posed).
The effect of finite $B$ is most important
near isolated zeros and singularities  of $z_\zeta(\zeta,0)$, 
where a regular perturbation
expansion in $B$ breaks down. (Away from these points the 
perturbation expansion is regular, at least initially.)
For the class of solutions~(\ref{fepsilon}) we are discussing, 
the isolated singularities of $z_{\zeta}(\zeta,0)$ are simple poles.
The theory suggests that the  introduction of finite surface tension
modifies the poles  ($\zeta_s$) by transforming  them  
into localized clusters of $-4/3$ singularities, but these
clusters move at leading order  according to the $B=0$ dynamics.  
Thus the  effect  of one
of these clusters on the interface is approximately 
equivalent to  that of the unperturbed ($B=0$) pole-like
singularity that has given birth to it. 

The influence of 
surface tension on the zeros of $z_\zeta(\zeta,0)$ is more complex.  
Each initial
zero  instantly gives birth to two localized inner regions, i.e., 
regions where the
$B=0$ and $B>0$ solutions differ by $O(1)$. One of the two
inner regions  moves, at least initially,  according to the $B=0$ dynamics 
of the original zero $\zeta_0$~\footnote{This region can move 
differently from the $B=0$ zero
once the  second inner region discussed below has impacted 
the unit disk.  We do not
consider this possibility here.}.
 Since the particular
zero surface tension solutions considered here have zeros that 
are  either bounded away from the unit disk for all time  or impact the 
unit disk  only after long times, the inner
region around $\zeta_0(t)$ has a negligible influence on the
interface.
The second inner region  created around
$\zeta_0(0)$ moves differently.
 The theory suggests that
this inner region consists of a cluster of singularities, whose size
scales like $B^{1/3}$.  Unlike the case
discussed above  this second inner region
moves away from the $B=0$ zero since, to leading order
in $B$, it moves   like a  {\it singularity} of the zero 
surface tension  problem and this speed
is different form the speed of the  zero $\zeta_0(t)$ which 
spawned the cluster.  As this singularity cluster approaches
the physical domain it may perturb the flow and the interface shape  may 
differ significantly from that at  $B=0$ shape.  The location of this
singularity cluster will be denoted  by $\zeta_d(t)$, and 
following \cite{Tanveer93}
we shall call it the daughter singularity. We emphasize that the 
dynamics of the daughter
singularity  cluster
is determined at lowest order solely by the $B=0$ solution $z^0(\zeta,t)$, 
at least
until  it arrives at the surroundings of the unit circle, and therefore
can be simply computed once the initial locations of the zeros 
of $z_\zeta (\zeta,0)$ are determined.

The daughter singularity 
evolution equation is given by (see  \cite{Tanveer93})
\begin{equation}
\label{evdaugh}
\dot{\zeta}_d(t)=-q_1^0(\zeta_d(t),t);\;\;\zeta_d(0)=\zeta_0(0)
\end{equation}
where \(q_1^0\) is defined by
\begin{equation}
\label{defq1}
q_1^0=\frac{\zeta}{2\pi i}\oint_{|\zeta'|=1}\frac{d\zeta'}{\zeta'}
\frac{\zeta+\zeta'}{\zeta'-\zeta}\frac{{\rm Re}\left[\zeta'W^0_\zeta
(\zeta',t)\right]}{|z^0_\zeta(\zeta',t)|^2}
\end{equation}
and the superscript 0 denotes that the  function evaluations are done using the
corresponding $B=0$ solution. The function $-q_1^0(\zeta,t)$ also gives the
characteristic velocity of a pole or branch point singularity 
of $z_\zeta (\zeta,t)$
 located at position $\zeta$ in the region
$|\zeta|>1$.
The initial position $\zeta_d(0)$  is a consequence of the fact that 
each  zero $\zeta_0(0)$ of the
zero surface tension solution will give birth to a daughter
singularity.
From Eq.~(\ref{evdaugh}) it can be shown \cite{Tanveer93} that 
$d |\zeta_d|/dt < 0$, so that
the daughter singularity approaches the unit circle and it  can
impact it in a finite time $t_d$, the daughter singularity impact 
time, satisfyng $|\zeta_d(t_d)|=1$. In the limit $B\rightarrow 0$, the daughter
singularity impact time  $t_d$ 
signals the time when the effects of the surface tension are felt on
the physical interface. For times larger than $t_d$ the $B=0$ interface
and the $B\rightarrow 0$ are expected to  differ significantly.

For the family of exact $B=0$ solutions  the mapping function~(\ref{fepsilon})
has four pole-like singularities: $\pm\zeta_s$ 
and $\pm\overline{\zeta}_s$, and 
four zeros $\pm\zeta_{0+}$ and $\pm\zeta_{0-}$ of $z_\zeta$ located at
\begin{widetext}
\begin{subequations}
\label{location}
\begin{eqnarray}
\label{location1}
\zeta_{0+}^2=\frac{-(\lambda+i\epsilon)\zeta_s^2-(\lambda-i\epsilon)
\overline{\zeta}_s^2}{2(1-2\lambda)}+ 
 \frac{\sqrt{\left[(\lambda+i\epsilon)\zeta_s^2+
(\lambda-i\epsilon)\overline{\zeta}_s^2\right]^2+4(1-2\lambda)
|\zeta_s|^4}}{2(1-2\lambda)} 
\\
\label{location2}
\zeta_{0-}^2=\frac{-(\lambda+i\epsilon)\zeta_s^2-(\lambda-i\epsilon)
\overline{\zeta}_s^2}{2(1-2\lambda)}- 
\frac{\sqrt{\left[(\lambda+i\epsilon)\zeta_s^2+
(\lambda-i\epsilon)\overline{\zeta}_s^2\right]^2+4(1-2\lambda)
|\zeta_s|^4}}{2(1-2\lambda)}. 
\end{eqnarray}
\end{subequations}
\end{widetext}
For the particular case $\lambda=1/2$ this solution 
presents only one pair of zeros $\pm\zeta_0$ located at
\begin{equation}
\label{location_half}
\zeta_{0}^2=\frac{|\zeta_s|^4}{2[(\lambda+i\epsilon)\zeta_s^2+
(\lambda-i\epsilon)\overline{\zeta}_s^2]}.
\end{equation}
In the following it will be useful to define the real  quantity
$\beta=-(\lambda+i\epsilon)\zeta_s^2-(\lambda-i\epsilon)
\overline{\zeta}_s^2$ which appears in
(\ref{location}) and (\ref{location_half}). 

Depending on the value
of $\lambda$ the initial data may have zeros on both  the real
and imaginary axes, or  all the zeros may lie on a single axis.
This difference has 
significant consequences in  the finite surface tension dynamics.
More specifically, when $\lambda< 1/2$ the zeros described in 
(\ref{location1}) and (\ref{location2})
are located on both the real and imaginary axes of $|\zeta|>1$, namely
at $\pm |\zeta_{0+}|$ and $\pm i |\zeta_{0-}|$. 
The situation is different for $\lambda>1/2$, which is further divided into
two cases, depending on whether $\beta^2+4 (1-2 \lambda)|\zeta_s|^2>0$
or $<0$. In the former case all four singularities lie on the real axis
(for $\beta>0$) or  on the imaginary axis (for $\beta<0$).  In the latter case
the four zeros are located off the axes in conjugate pairs, i.e.
at $\pm \zeta_0$ and $\pm \bar{\zeta}_0$.
Finally, when $\lambda=1/2$ the
solution~(\ref{fepsilon})  has only  two zeros, located on the real 
axis at $\pm |\zeta_s|^2/\sqrt{-2 \beta}$
when $\beta<0$ and on the imaginary axis at $\pm |\zeta_s|^2/\sqrt{2 \beta}$
when $\beta>0$.
Note that for  $\lambda=1/2$ the $B=0$ solution has two less  zeros
than for  $\lambda\neq1/2$. 

The initial  zero locations  described above have a critical bearing
on whether the daughter singularity will impact the unit 
disk~\footnote{Although the daughter singularity
is said to impact the unit disk when  $|\zeta_d(t)| =1$, the  singularities
comprising the cluster do not actually impinge upon the unit disk.  However, 
at $t_d$ the singularities are close enough to influence 
the interface shape, in the sense
that $|z_\zeta(\zeta_d,t_d)-z^0_\zeta (\zeta_d,t_d)|=O(1)$.}. Although
all daughter singularities approach the unit disk, their impact may be shielded
by the presence of  an inner region corresponding to 
a pole singularity.  More precisely, since
$\zeta_d$ and $\zeta_s$ obey the same dynamical equation, they will move
together
if they get close enough to each other.  
However, the inner region around a pole
moves to leading order like the $B=0$ pole, i.e., 
it moves exponentially slowly toward $|\zeta|=1$ 
when $|\zeta_s|-1<<1$, and does not impinge upon the unit 
disk in finite time \cite{Howison86}.   In this case
the $O(B^{1/3})$ inner region around the daughter singularity will not
affect the dynamics on $|\zeta|=1$, at least until $t=O(-\ln B)$.  
Before this time, we
expect the interface to be uninfluenced by the presence of 
the daughter singularity.
This shielding mechanism is discussed in the context of single fingers in 
\cite{Siegel96b}.

\begin{figure}
\includegraphics{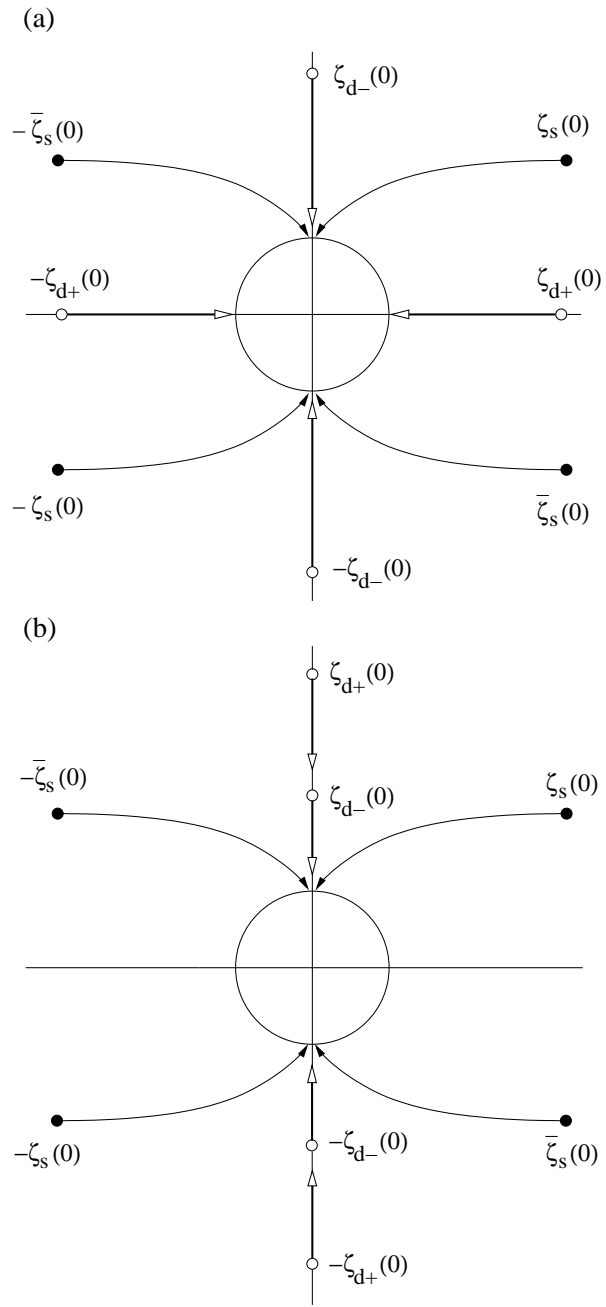}
\caption{\label{fig:scheme}
(a) Schematic representation of the dynamics of pole ($\zeta_s$) and 
daughter ($\zeta_d$) singularities for $\lambda<1/2$.
(b) Schematic representation of one of the two possible dynamics of 
pole ($\zeta_s$) and
daughter ($\zeta_d$) singularities for $\lambda>1/2$.}
\end{figure}

Knowledge of the $t \rightarrow \infty$ asymptotic state and 
the initial locations
of zeros  can be used to 
ascertain whether shielding can occur.  The $B=0$ asymptotic state
corresponds to $\zeta^2_s (t \rightarrow \infty) \rightarrow  \pm 1$.
Thus, for $\lambda<1/2$, only one pair of daughter 
singularities may be shielded---never both---
so at least one pair of daughter singularities will impinge on the unit disk.
The daughter singularities will also not be shielded when 
$\lambda>1/2$ and  $\beta^2+4 (1-2 \lambda)|\zeta_s|^2<0$.
However, for $\lambda>1/2$ and  $\beta^2+4 (1-2 \lambda)|\zeta_s|^2>0$ 
it is possible
for all the daughter singularities to be shielded, 
since they lie on a single axis.
The daughter singularities can also be completely shielded when $\lambda=1/2$.
The different possibilities are schematically depicted 
in Fig.~\ref{fig:scheme}.

\begin{figure}
\includegraphics{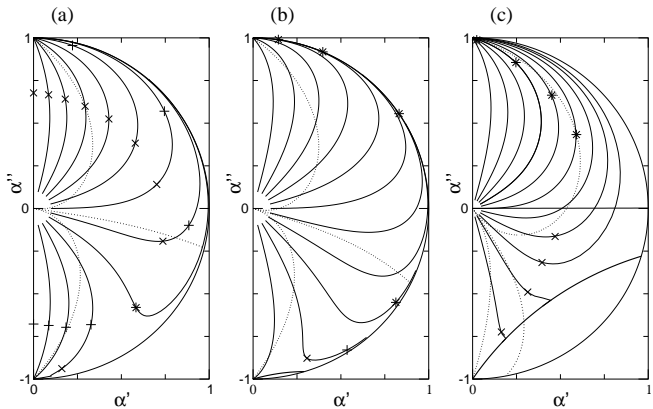}
\caption{\label{fig:daughters}Phase portraits for (a) $\lambda=1/3$ 
and $\epsilon=0.1$, 
(b) $\lambda=2/3$ and $\epsilon=0.1$, and (c) $\lambda=1/3$ and $\epsilon=1/2$.
The daughter singularity impact is indicated by the symbols. 
The $+$ symbol corresponds 
to the impact of $\zeta_{d+}$, $\times$ to the impact of $\zeta_{d-}$ 
and $*$ to
the simultaneous impact of $\zeta_{d+}$ and $\zeta_{d-}$.}
\end{figure}

We have numerically computed the daughter singularity impact time $t_d$
for various values of $\lambda$ and $\epsilon$, using initial conditions
close to the planar interface, $|\zeta_s|^2=20$ and various values
of ${\rm Arg}[\zeta_s^2]$. 
Figure~\ref{fig:daughters}  shows the phase portrait for 
different values of $\lambda$ 
and $\epsilon$ with the daughter singularity impact indicated. 
From the plots it is immediately seen that for $\lambda < 1/2$ 
at least one daughter singularity always hits the unit circle, and for 
$\lambda \geq 1/2$ some trajectories are free from daughter singularity 
impact. In addition, it is observed that for fixed $\lambda$ a larger
value of $\epsilon$ causes the daughter singularities to hit in shorter
times (or less developed fingers) than a smaller value of $\epsilon$, 
and for fixed $\epsilon$ larger $\lambda$ implies larger impact times.
We have also checked that the daughter singularity impact occurs well
before a finite time singularity, i.e., the impact of a zero of $z_\zeta$. 
Thus, the
effect of surface tension is significant well  before the curvature in the
zero surface tension solution becomes large.

It is noted that the $\lambda$ dependence of the daughter singularity 
impact  is consistent with
the results of steady state selection theory 
\cite{Shraiman86,Hong86,Combescot86,Tanveer87}. 
According to selection theory,
for small $B$ the possible values of $\lambda$ are discretized: $\lambda$
must satisfy the relation $\lambda=\lambda_n(B)$, given to leading order
by
\begin{equation}
\label{selection}
\lambda_n(B)=\frac{1}{2}\left\{1+(\frac{1}{8}\pi^2C_nB)^{2/3}\right\},
\:\:n=0,1,2,...
\end{equation}
where $n$ parameterizes the branch of solutions. Note that 
$\lambda_n>1/2$ for all $n$.  
The steady finger shape is  to leading order a Saffman-Taylor finger, 
with the above
values of $\lambda_n$ substituted for the width $\lambda$. 
On the other hand for $\epsilon>0$ the asymptotic state of~(\ref{fepsilon}) 
is a Saffman-Taylor
finger of width $\lambda$. From Eq.~(\ref{selection}) it is clear
  there exists a steady solution with width $\lambda_n(B)$ close to 
a Saffman-Taylor  finger of arbitrary width $\lambda>1/2$.  Thus the 
shielding of the daughter singularity,
which leads to the persistence of a Saffman-Taylor solution with 
$\lambda>1/2$ over long
times, is consistent with  steady state selection theory~\footnote{The
steady state with $\lambda>1/2$ is unstable to tip splitting modes, 
although  specifying
the initial value problem in the extended complex plane precludes 
the presence of noise needed
to activate the instability.}.  In contrast for $\lambda<1/2$ there are 
no nearby steady
solutions.  Thus, a Saffman Taylor finger with $\lambda<1/2$ cannot 
persist over a long time.
We see that the impact of a daughter singularity provides a mechanism for 
the onset of finger competition,
finger widening, and selection of a width $\lambda>1/2$.

For $\epsilon=0$ the scenario is similar, except there is an added class of
exact $B>0$  solutions. Magdaleno and Casademunt 
\cite{Magdaleno99} have shown 
that two-finger solutions composed of steadily propagating but unequal fingers
do exist  for small nonzero $B$.
The introduction of a small nonzero surface tension selects
a discrete set of solutions from the continuum of fixed points 
of the $B=0$ phase portrait.  The solutions are 
parameterized by the total width of the fingers 
$\lambda=\lambda_1+\lambda_2$ and the relative width $q=\lambda_1/\lambda$,
and the introduction of finite $B$ discretizes the possible values of the
parameters.  In particular, they must satisfy a condition of the form 
$\lambda=\lambda_n(B)$ 
and $q=q_{n,m}(B)$
where $n$ and $m$ are integers. The expression for $\lambda_n(B)$ at 
lowest order is 
equivalent to Eq.~(\ref{selection}), but with different coefficients $C_n$.
The shape of these solutions are given to leading order (in the limit 
$t \rightarrow \infty$) by
(\ref{fepsilon}) with allowed value of $\lambda_n(B)$ substituted for 
the width $\lambda$.
Again, $\lambda_n(B)>1/2$, and the consistency between  daughter singularity
impacts and steady state selection theory follows as above.

We conjecture that the  outcome of interfacial shape evolution after 
the daughter
singularity impinges  is in general independent of the particular finger 
on which the impact first occurs
i.e., independent of the point  at
which $\zeta_d(t)$ impacts on $|\zeta|=1$.   More specifically, 
we surmise that 
impact  on either the shorter (trailing) or larger (leading) finger  
retards the velocity of that finger, and is accompanied
by the widening of the leading  finger, so as to maintain a constant 
fluid flux at infinity.
The widened leading finger then shields the trailing finger, preventing 
it from further
growth.  Thus, the finger which is leading at the time of the 
daughter singularity
impact `wins' the competition,  in the sense that it will evolve for
$t \rightarrow \infty$ to the ST steady finger.
To  examine this conjecture and study the dynamics of finger competition 
with finite (but small)
surface tension we have numerically computed the evolution of an 
interface with initial conditions given by the conformal 
mapping Eq.~(\ref{fepsilon})
close to the planar interface ($|\zeta_s(0)|^{-2} \ll 1$).  The results
are reported in the next section. 

\section{Numerical Results}
\label{numerics}

Numerical computations  have been performed for $B>0$,  using an
initial interface corresponding to 
the explicit $B=0$ solutions discussed in Sec.~\ref{zero_b}.
The effect of positive surface tension  on this class of solutions
is explored for
 various values of $\epsilon$ and a variety of initial pole positions.

We employ the numerical method introduced by Hou \it et al. \rm
\cite{Hou94} and used in other studies of  small surface tension
effects in Hele-Shaw flow   
\cite{Siegel96a,Siegel96b,Ceniceros99,Ceniceros00}. The method
is described in detail in Ref.~\cite{Hou94}. It is a 
boundary integral method in which the interface is parameterized at
equally spaced points by means of an equal-arclength variable $\alpha$. 
Thus, if $s(\alpha,t)$ measures arclength along the interface then
the quantity $s_{\alpha}(\alpha,t)$ is independent of $\alpha$ and
depends only on time. The interface is described using the tangent
angle $\theta(\alpha,t)$ and the interface length $L(t)$, and these
are the dynamical variables instead of the interface $x$ and $y$ 
positions. 
 The evolution equations are written in
terms of $\theta(\alpha,t)$ and $L(t)$ in such a way that the 
high-order terms, which  are responsible of
the numerical stiffness of the equations,  appear linearly
and with constant coefficients.
This fact is exploited in the construction of an efficient
numerical method, i.e., one that has no time step constraint
associated with the surface tension term yet is explicit in Fourier space.
We have used a linear 
propagator method that is second order in time, combined with a 
spectrally accurate
spatial discretization.  Results in this section are specified in terms of the 
scaled variables
\begin{equation}
\tilde{t}=\pi t; \:\:\:\tilde{B}=\pi^2B; \:\:\: \tilde{x}=\pi x; 
\:\:\: \tilde{y}=\pi y.
\end{equation}
instead of the original ones used in previous sections.

The number of discretization
points is chosen so that all Fourier modes of $\theta(\alpha,t)$ with 
amplitude greater than round-off are well resolved, and as soon as the
amplitude of the highest-wavenumber mode becomes larger than the filter
level the number of modes is increased, with the amplitude of the additional
modes initially set to zero. The time step $\Delta t$ is decreased until an 
additional decrease does not change the solution to plotting accuracy, 
nor lead to any significant differences in any quantities of interest.
In a typical calculation 512 discretization points are initially used, 
and the initial time step is $\Delta t=5\cdot10^{-4}$.
For small values of surface tension numerical noise is a major problem, and
the spurious growth of short-wavelength modes induced by round-off error must
be controlled. To  help prevent this noise-induced growth at short wavelengths
spectral filtering \cite{Krasny86} is applied. Additionally,
we minimize noise effects and also  assess the time at which these 
effects become prevalent
by employing extended precision calculations, as described in the 
next section.  
\begin{figure*}
\includegraphics{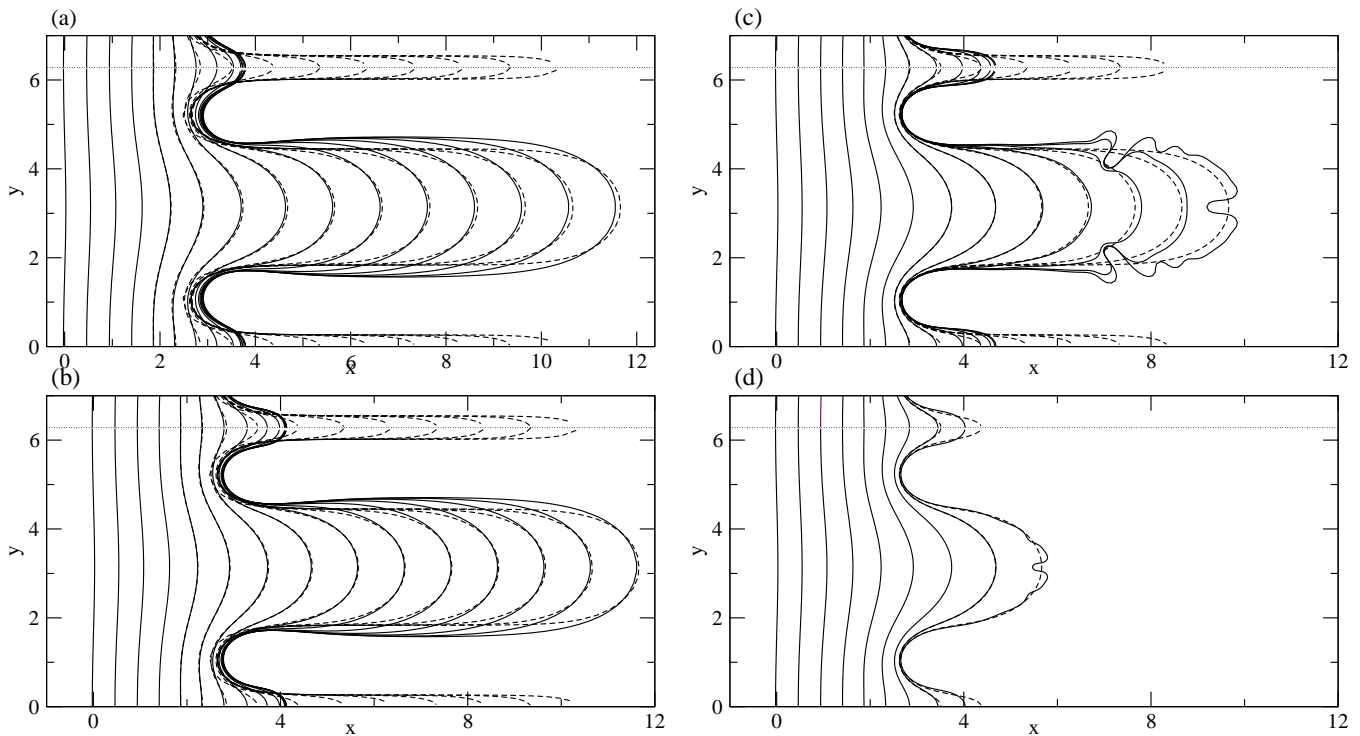}
\caption{\label{fig:ep0} Evolution of an initial condition of the 
form~(\ref{fepsilon}) with
$\lambda=1/2$,  $\epsilon=0$ and $\zeta^2_s(0)=20\exp(i\pi/6)$. The solid lines
correspond to surface tension $\tilde{B}$ 
values (a) 0.01, (b) 0.005, (c) 0.001 and
(d) 0.0005. The dashed lines correspond to the zero surface tension evolution.
The time difference between different curves is 0.5. The physical channel 
in the $y$ direction extends from the origin to the dotted line, and the 
region 
above is plotted for better visualization of the lateral finger.}
\end{figure*}

Our main interest is to uncover the role of surface tension in the dynamics
of finger competition. To isolate the features of finger competition from those
of width selection, 
we will concentrate on $B=0$ solutions with $\lambda=1/2$, the value 
selected by surface tension in the limit $B\rightarrow 0$. Since the $B=0$ dynamics for $\epsilon=0$ and $\epsilon\neq0$ is quite different
the numerical results for the two cases will be presented separately.

\subsection{Solutions with $\epsilon = 0$}
We first consider parameter values $\lambda=1/2$ and $\epsilon=0$.
A typical set of interfacial profiles is shown in Fig.~\ref{fig:ep0}. 
The initial data  is given by the mapping function Eq.~(\ref{fepsilon}), with $\lambda=1/2$,
$\epsilon=0$, $d(0)=0$ and $\zeta^2_s(0)=20\exp(i\pi/6)$. With this value 
of $\zeta^2_s(0)$ the initial interface is well inside the linear regime.
Evolutions are shown for different values of $\tilde{B}$, and the $B=0$ interface
evolution is also plotted for comparison. In all these evolutions the filter
level is set to $10^{-13}$, although later we shall make comparisons
to profiles computed at higher precision.

For the largest value of surface tension the computed $B>0$ and the exact 
$B=0$ solutions first differ appreciably at  the seventh curve, 
corresponding to $\tilde{t}\approx 3$. At this  point  the velocity of the
small finger (at the channel sides)
begins to decrease
 and  it is clearly left behind when compared with the small
finger evolution in the  $B=0$ solution. Eventually, the advance of the 
small finger is completely suppressed and the larger finger widens to
attain a width close to $1/2$ of the channel. For a smaller value of 
surface tension, for instance $\tilde{B}=0.001$, the evolution displays qualitatively 
the same behavior.  The $B>0$ interface differs appreciably from the $B=0$ sightly
later than before (i.e., at the eighth curve) and the 
region where the two solutions differ most  
is to some extent more localized around the small finger than
for larger values of $B$. Additionally, for this value of surface tension the effect of
numerical noise is   
clearly exhibited in the interfacial profiles. Here the tip-splitting and 
side-branching activities are a clear effect of numerical noise, as can be easily checked
redoing the computation with a different noise filter level.

\begin{figure}
\includegraphics{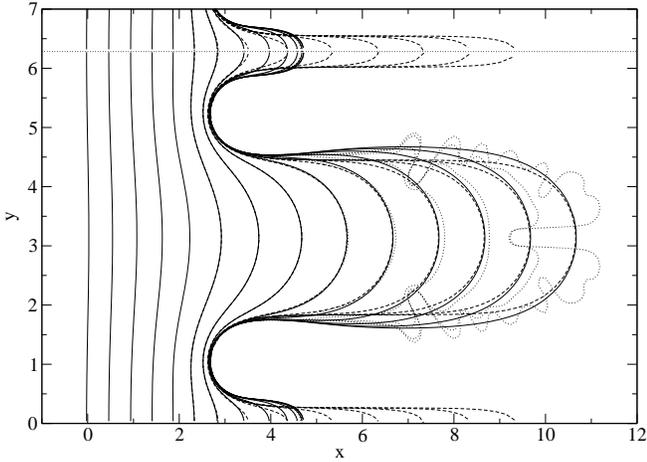}
\caption{\label{fig:noise}  Evolution of an initial condition of the form~(\ref{fepsilon}) with
$\lambda=1/2$,  $\epsilon=0$ and $\zeta^2_s(0)=20\exp(i\pi/6)$. The solid line
correspond to $\tilde{B}=0.001$ with a filter level equal to $10^{-27}$,
the dotted line corresponds to the same $\tilde{B}$ but with the filter level equal
to $10^{-13}$ and the dashed line corresponds to the zero surface tension
solution. The time difference between curves is 0.5.
As in Fig.~\ref{fig:ep0}, the physical channel
in the $y$ direction extends from the origin to the dotted line.}
\end{figure}

 In order to suppress
or delay the branching induced by numerical noise that appears for small values
of surface tension it is necessary to use higher
precision arithmetic, e.g. quadruple precision  (128-bit arithmetic).  The filter
level can then  be reduced by a large amount and the outcome of spurious oscillations is 
substantially delayed.  Figure~\ref{fig:noise} shows the effect
of reducing the filter level  to
$10^{-27}$. The $B=0$ solution is plotted, as well as the computation with double precision.
  For $\tilde{B}=0.001$ the branching is totally suppressed, at least for
the times we have computed, but for smaller values of $\tilde{B}$ the use of quadruple 
precision is only able to delay the branching and not totally   suppress it. The quadruple precision computation confirms the results observed with lower
precision: the introduction of finite (but small) surface tension results in the
suppression of the small finger. From Fig.~\ref{fig:noise} one can also see 
that for long times,
when the interface is clearly affected by numerical noise (in the double precision curve),
the noise-induced branching is restricted to the large finger, and the small finger
is basically unaffected by noise. This observation suggests that the small finger shape,
as well as  its tip velocity and tip curvature, can be trusted even when the large finger
has developed tip-splittings and side-branchings due to the spurious growth of 
round-off error.

\begin{figure}
\includegraphics{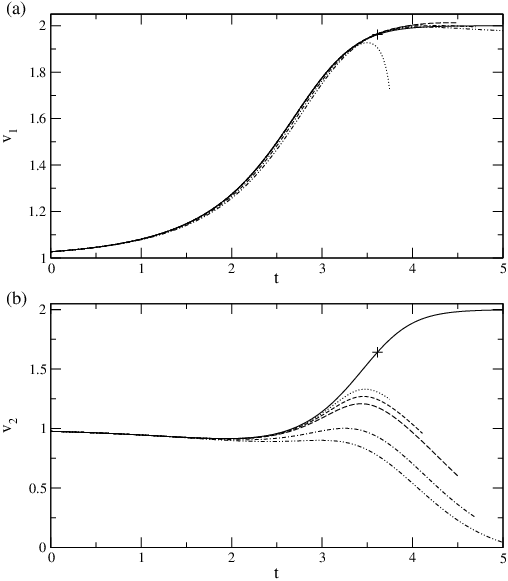}
\caption{\label{fig:tip0} Computed tip velocities for the initial condition of 
Fig.~\ref{fig:noise} , (a)
corresponds to the central (large) finger and (b) to the lateral (small) finger.
The daughter singularity impact time $\tilde{t}_d$ is indicated by the $+$ symbol. The value
of $\tilde{B}$ is: 0 (solid line), 0.0002 (dotted line), 0.0005 (dashed line),
0.001 (long dashed line), 0.005 (dot-dashed line) and 0.01 (dot-dot-dahed line).}
\end{figure}

Figure~\ref{fig:tip0}  shows the tip velocity of both 
fingers versus $\tilde{t}$ for decreasing values of
surface tension. It can be seen that the velocity of the large finger is  
only slightly
affected by surface tension, whereas the velocity of the small finger
is substantially reduced by the inclusion of finite $B$. As $\tilde{B}$ is decreased the
tip velocity of the small finger is more faithful to the $B=0$
evolution before the daughter singularity impact (shown by a cross), and clearly veers away from the $B=0$ velocity later in the 
evolution, consistent with asymptotic theory. Note that at the smallest value of $\tilde{B}$  the tip
velocity of the large finger 
drastically differs from the $B=0$ velocity  at late times. This
discrepancy  is a manifestation of  noise effects
in the neighborhood of the large finger tip. 
However, as previously seen, the 
small finger is basically unaffected by noise at the times we have plotted.

\begin{figure}
\includegraphics{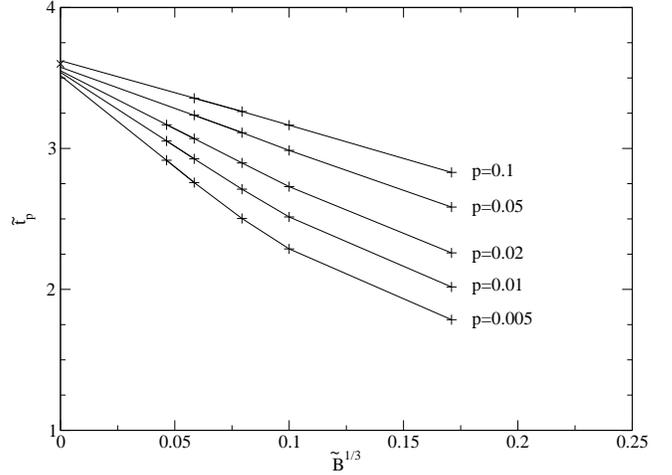}
\caption{\label{fig:scaling} The time $\tilde{t}_p$ (defined in the text) 
versus $\tilde{B}^{1/3}$. From top to
bottom, p=0.1, 0.05, 0.02, 0.01, 0.005. The daughter singularity impact time $\tilde{t}_d$
is inducated by a $\times$ symbol, and the curves are linearly extrapolated
for comparison.}
\end{figure}

In order to further verify that the daughter singularity impact is responsible
for the observed change in the small finger tip speed we follow the scheme 
introduced in \cite{Siegel96b}.
Define $t_p$ as the time when the computed tip velocity differs by $p$ from the
$B=0$ tip velocity. According to asymptotic theory this $t_p$ will be a linear function of $B^{1/3}$ in the limit
$B \rightarrow 0$ as long as $p$ is small enough. 
Figure~\ref{fig:scaling} shows $\tilde{t}_p$ versus $\tilde{B}^{1/3}$
for various values of $p$, and it can be seen that $\tilde{t}_p$ exhibits the predicted 
behavior. Moreover, we have extrapolated the $B=0$ value of $\tilde{t}_p$ using the two
points of lowest $\tilde{B}$ and the result is very close to $\tilde{t}_d$, 
whose value is represented
by a cross. We conclude that the impact of the daughter singularity
is associated with the dramatic change of the $B>0$ solution when
compared to the zero surface tension solution, reducing the  velocity of the small
finger and eventually suppressing it.
In contrast, for the $B=0$ dynamics the small 
finger `survives', propagating with the  same asymptotic speed as the larger finger.
Note that
the average interface advances at unit velocity, and a tip velocity below $1$ implies
that the finger is \it retreating \rm in the reference frame of the average interface.

In summary then, our numerical results show that the computed interface for $B \neq 0$  follows
 the $B=0$  evolution  for
an $O(1)$ time interval---roughly corresponding to the daughter singularity impact time---and
that
at further times   the velocity
of the small finger decreases while
 the large finger widens.  The small finger eventually  comes to a halt and the larger (leading) finger reaches an
asymptotic width slightly above $1/2$,
the width singled out by  selection theory.
It is noted that
for the initial condition we have studied
the daughter singularity impact takes place on the
tip of the small finger. Therefore, the influence of  surface tension on the
interface should be significant first around the impact point, that is,
the small finger tip. Our numerical results show that in fact this is the case;  the  initial
effect of the daughter
singularity impact is to slow and then completely
 stop the growth of the small finger. Later on, as the
singularity cluster centered in $\zeta_d$ spreads over the unit circle,
the effect of surface tension is felt by the whole interface and the large finger
widens to reach the selected width.

We have also studied the finite surface tension dynamics for
a more general class of initial 
conditions. More precisely, we have
studied initial conditions of the form $\zeta^2_s(0)=20\exp(i\;n\pi/12)$ where
$n=0, \pm 1,..., \pm 6$, and have obtained the same qualitative results 
as  in the case previously studied, namely that the presence of 
small  surface tension suppresses the growth
of the finger which is trailing at the time of daughter singularity impact.
In order to compare the $B=0$ and the $B \neq 0$
dynamics in a compact and global way we have plotted the phase portrait for $B=0$ using the 
the tip velocities $v_1, \,v_2$ as dynamical variables. In the laboratory frame
they read
\begin{subequations}
\label{velocitats}
\begin{eqnarray}
v_1=\frac{1+{\rm i}(\zeta^2_s-\bar{\zeta}^2_s)+\zeta^2_s\bar{\zeta}^2_s}
    {\zeta^2_s\bar{\zeta}^2_s+{\rm i}(\zeta^2_s-\bar{\zeta}^2_s)/2}
\\
v_2=\frac{1-{\rm i}(\zeta^2_s-\bar{\zeta}^2_s)+\zeta^2_s\bar{\zeta}^2_s}
    {\zeta^2_s\bar{\zeta}^2_s-{\rm i}(\zeta^2_s-\bar{\zeta}^2_s)/2}.
\end{eqnarray}
\end{subequations}
Now a comparison between dynamics  for
$B=0$ and  $B \neq 0$ is straightforward since  the trajectories can be plotted together and compared. In 
addition, the tip velocity is a useful variable because it contains
geometric information; specifically the inverse of the tip velocity is equal to the width of the
finger in the asymptotic ($t \rightarrow \infty$) regime. It is important to note that $(v_1, v_2)$
are dynamical variables for the $B=0$ problem, so that  the plot of 
the zero surface tension trajectories onto the space $(v_1, v_2)$ is a true 
phase portrait.  On the other hand  $(v_1, v_2)$ are not state variables of
the problem with finite surface tension, so in this case we simply
obtain  a projection onto
 the $(v_1, v_2)$ space
of the original $B\neq 0$ trajectory, which is embeded in the infinite-dimensional 
phase space of interface configurations. 

\begin{figure}
\includegraphics{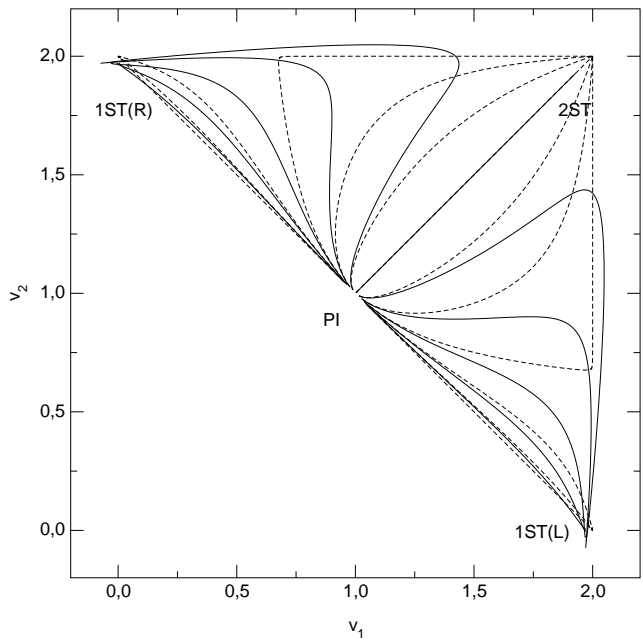}
\caption{\label{fig:vel0} Plot of the evolution of initial conditions of 
the form~(\ref{fepsilon}) with
$\lambda=1/2$,  $\epsilon=0$ and $\zeta^2_s(0)=20\exp(i\;n\pi/12)$ and
$n=0, \pm 1,..., \pm 6$ in the $(v_1,v_2)$ or tip speed space. The solid
line corresponds to $\tilde{B}=0.01$ and the dashed line to $\tilde{B}=0$.}
\end{figure}

Figure~\ref{fig:vel0} shows the phase portrait for $B=0$ together with the tip velocities
obtained from the initial conditions described above for $\tilde{B}=0.01$.
From the figure it is evident that the introduction of 
 finite surface tension has substantially changed the global phase dynamics of the problem.
Only one $\tilde{B}=0.01$ trajectory connects the planar interface $(1,1)$ and the 2ST point 
$(2,2)$, corresponding to the unsteady double Saffman-Taylor finger.
Any other
$\tilde{B}=0.01$ trajectory ends in one of the two ST finger points, ST(L) at $(2,0)$ and ST(R) at $(0,2)$. In
contrast, the $(2,2)$ point,  equivalent to the continuum of fixed points 
present with the $(\alpha ', \alpha'')$ or $({\rm Re}\zeta_s,{\rm Im}\zeta_s)$ 
variables, has a finite basin of attraction for $B=0$.
The introduction of finite surface tension has  dramatically changed the zero surface 
tension $(v_1, v_2)$ trajectories, to the extent that the $B=0$ phase portrait
and the $B\neq0$ projection are not topologically equivalent. 
This result
is not a complete surprise, since it was anticipated 
from the structural instability
of the dynamical system governing the evolution of Eq.~(\ref{fepsilon}) 
for $\epsilon=0$ \cite{Magdaleno98}.  A more dramatic example of  topological
inequivalence of phase portraits will be given in the next subsection, when
we consider the case $\epsilon \neq 0$.
\begin{figure}
\includegraphics{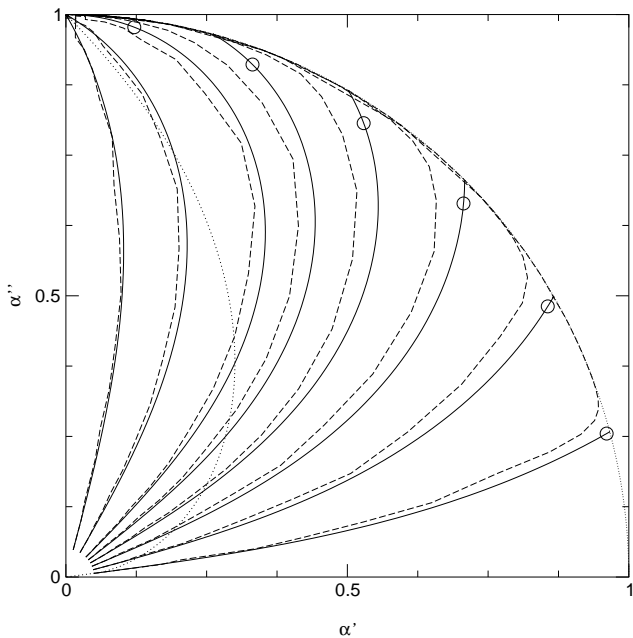}
\caption{\label{fig:proj0} Comparison between the $\tilde{B}=0$ trajectories and the projected
evolutions with $\tilde{B}=0.01$, for the initial conditions of Fig.~\ref{fig:vel0}.
The solid line corresponds to $\tilde{B}=0$ and the dashed lines to the projection
of the $\tilde{B}=0.01$ evolutions. The daughter singularity impacts are indicated
by a circle.}
\end{figure}

Although the use of the variables $(v_1, v_2)$ has allowed us to project the finite
surface dynamics onto the zero surface tension phase portrait this projection has
one major limitation: it only considers a local quantity, the tip velocity. 
We have also
considered a  projection that takes more global properties
of the interface into account.  Specifically, given a computed  $B\neq0$ 
solution for  an initial condition
of the form~(\ref{fepsilon}),
one can use a suitable norm to 
define a `distance' between the computed interface and the $B=0$ interface
obtained from the mapping function Eq.~(\ref{fepsilon}).
We choose this `distance' to be the
area enclosed between the two interfaces at a given time. 
Additionally, we define a projection of the $B\neq0$ interface  onto the $B=0$ phase space (with phase space
variables $({\rm Re}\,\zeta_s,{\rm Im}\,\zeta_s)$) by selecting the  value of  $\zeta_s$ that minimizes
the `distance' between the two interfaces, with
the restriction that the position of the two mean interfaces
must be the same. The latter condition ensures that the projection satisfies mass conservation.

Figure~\ref{fig:proj0} shows the $B=0$ phase portrait and the corresponding projected evolution
for surface tension $\tilde{B}=0.01$. Again, the plot clearly shows that the 
introduction of finite surface tension modifies the phase
portrait of $B=0$. The projected trajectories are initially close to the $B=0$
dynamics, but for well developed fingers (corresponding to $|\alpha|\sim 1$)
the projection departs from the $B=0$ trajectory towards the Saffman-Taylor
fixed point, located at $\alpha'=0,\alpha''=1$. The projected trajectory only remains close to
the corresponding $B=0$ trajectory when the latter evolves towards the Saffman-Taylor
fixed point. More precisely, the continuum of fixed points present for $B=0$ has been removed
by surface tension and the Saffman-Taylor fixed point is the universal 
attractor of the dynamics 
for finite surface tension. 
\begin{figure}
\includegraphics{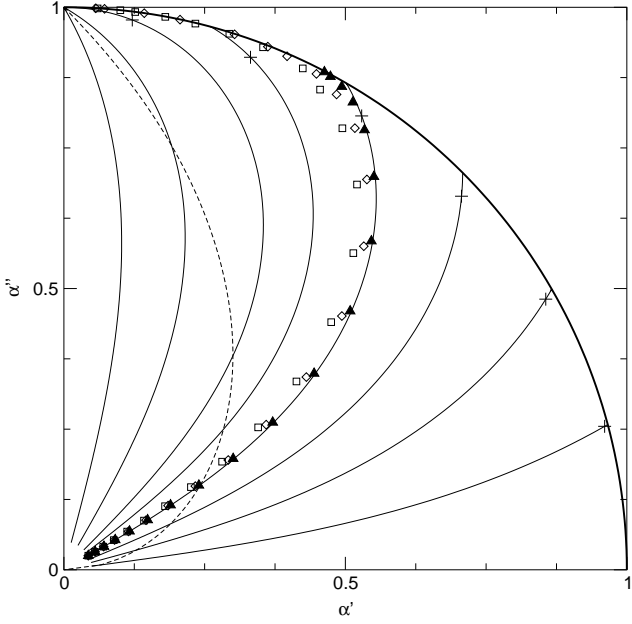}
\caption{\label{fig:proj0a} Comparison between the $\tilde{B}=0$ trajectories and the projection
of the evolution of the initial condition given by Eq.~(\ref{fepsilon}) with
$\lambda=1/2$, $\epsilon=0$ and $\zeta^2_s(0)=20\exp(i\pi/6)$, where $\blacktriangle$
corresponds to $\tilde{B}=0.001$, $\diamond$ to $\tilde{B}=0.005$, $\Box$ to
$\tilde{B}=0.01$ and $\times$ to $\tilde{B}=0$. The daughter singularity impacts
are indicated by a plus.}
\end{figure}

In Fig.~\ref{fig:proj0a} the projection for decreasing values 
of $\tilde{B}$ is plotted, using the 
initial condition $\zeta^2_s(0)=20\exp(i\pi/6)$. As $\tilde{B}$ is decreased the 
projected trajectory gets closer to the $B=0$ trajectory, but as it approaches the 
point when the daughter singularity impinges the unit circle (this point is 
signaled by a cross) the projection departs from the $B=0$ trajectory and
approaches  the Saffman-Taylor fixed point, consistent with
asymptotic theory.

\subsection{Solutions with $\epsilon \neq 0$}

The continuum of fixed points present for $\epsilon=0$ is absent 
for $\epsilon \neq 0$, but in this case finite-time singularities
in the form of zeros of $z_\zeta$ impinging on the unit disk do 
appear for some initial conditions. Therefore, we can expect that the 
effect of finite surface tension will be somewhat different than for
$\epsilon=0$. Firstly, the presence of  surface tension should eliminate finite-time
singularities, and secondly, finite $B$ could modify the basin of attraction
for the two attractors of the $B=0$ dynamical system, namely  the
side Saffman-Taylor finger and
the center Saffman-Taylor finger.

\begin{figure}
\includegraphics{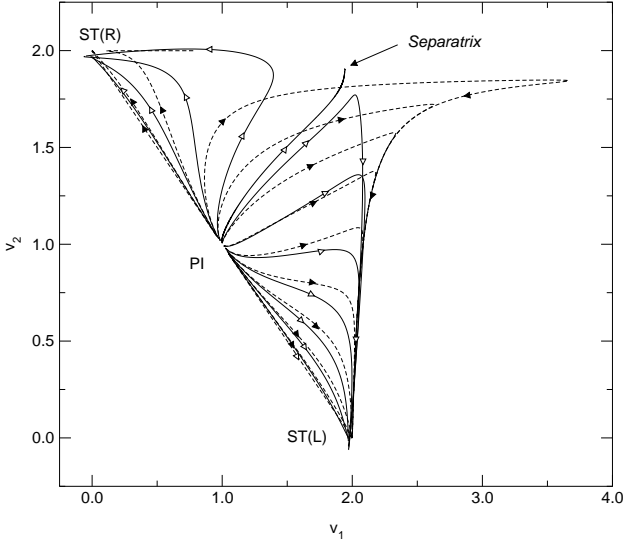}
\caption{\label{fig:vel01}  Plot of the evolution of initial conditions of the 
form~(\ref{fepsilon}) with
$\lambda=1/2$, $\epsilon=0.1$ and $\zeta^2_s(0)=20\exp(i\;n\pi/12)$ and
$n=0, \pm 1,..., \pm 6$ in the $(v_1,v_2)$ or tip speed space. The solid
line corresponds to $\tilde{B}=0.01$ and the dashed line to $\tilde{B}=0$.
The computed  trajectory that most nearly  separates the two basins of atraction is also plotted. Note
that the long time behavior of the third and fourth $\tilde{B}=0.01$ curves
(counting from the upper left trajectory in clockwise direction) is dramatically
different from the corresponding $\tilde{B}=0$ solutions.}
\end{figure}
\begin{figure*}
\includegraphics{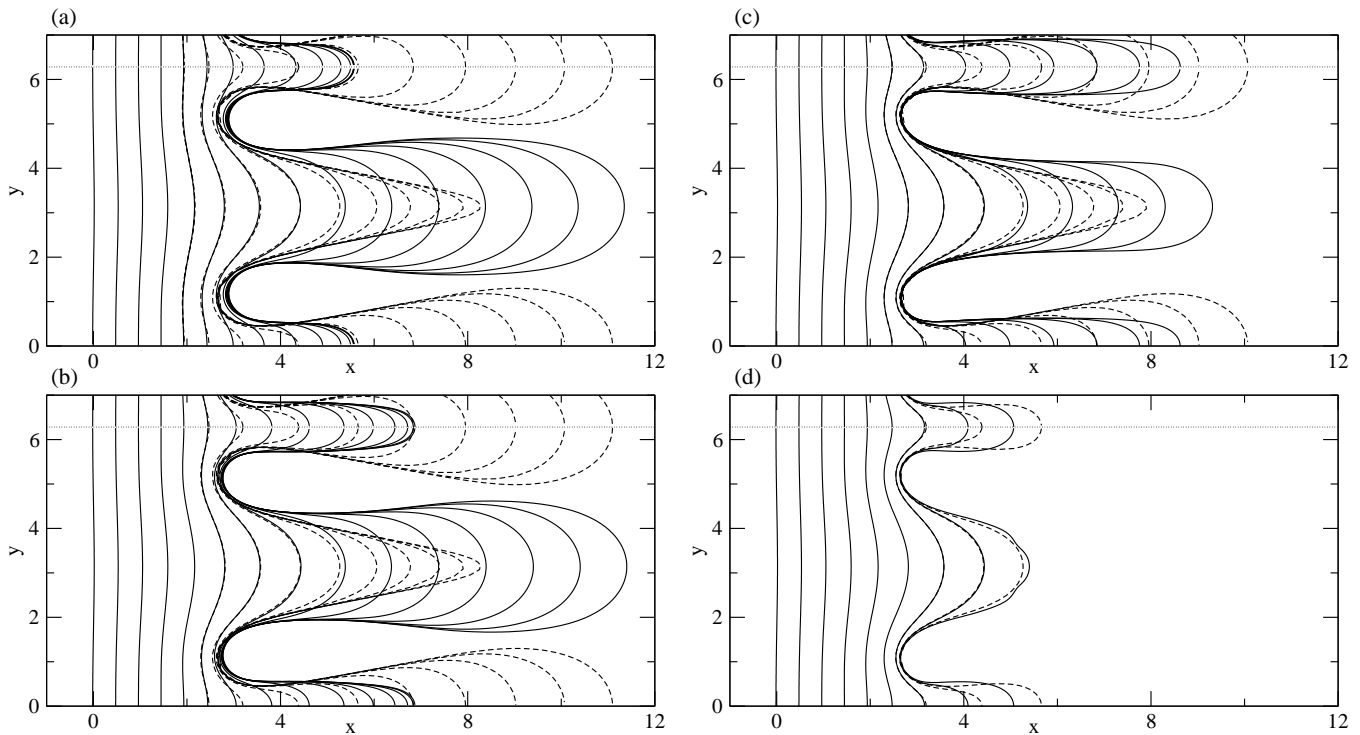}
\caption{\label{fig:ep01} Evolution of an initial condition of the form~(\ref{fepsilon}) with
$\lambda=1/2$,  $\epsilon=0.1$ and $\zeta^2_s(0)=20\exp(-i\pi/6)$. The solid lines
correspond to surface tension $\tilde{B}$ values (a) 0.01, (b) 0.005, (c) 0.001 and
(d) 0.0005. The dashed lines correspond to the zero surface tension evolution.
The time difference between different curves is 0.5.The physical channel
in the $y$ direction extends from the origin to the dotted line.}
\end{figure*}

To explore this,  we have performed computations with $\lambda =1/2$ 
and $\epsilon = 0.1$ with initial conditions $\zeta^2_s(0)=20\exp({\rm i}\;n\pi/12)$
 and $n=0, \pm 1,..., \pm 6$.  Initially we set $\tilde{B}=0.01$ and use
a value of the noise filter level
equal to $10^{-13}$, which suffices due to the relatively large value of $\tilde{B}$. 
The easiest
way to compare both dynamics, finite $B$ and $B=0$, is to plot their
trajectories in velocity space. Thus, in Fig.~\ref{fig:vel01}
the tip velocities $(v_1, v_2)$ of the $\tilde{B}=0.01$ computation are plotted together 
with the tip velocities for $B=0$. For arbitrary $\epsilon$ and 
$\lambda$ the tip velocities of the $B=0$ solution read 
\begin{subequations}
\label{+velocitats}
\begin{eqnarray}
v_1=\frac{1 + {\rm i}(\zeta^2_s - \bar{\zeta}^2_s) + \zeta^2_s\bar{\zeta}^2_s}
    {\zeta^2_s\bar{\zeta}^2_s - \epsilon(\zeta^2_s + \bar{\zeta}^2_s) 
    + {\rm i}\lambda(\zeta^2_s - \bar{\zeta}^2_s) - (1 - 2\lambda)}
\\
v_2=\frac{1-{\rm i}(\zeta^2_s-\bar{\zeta}^2_s)+\zeta^2_s\bar{\zeta}^2_s}
    {\zeta^2_s\bar{\zeta}^2_s + \epsilon(\zeta^2_s + \bar{\zeta}^2_s) 
    - {\rm i}\lambda(\zeta^2_s - \bar{\zeta}^2_s) - (1 - 2\lambda)}. 
\end{eqnarray}
\end{subequations}
From the plot one can see that most $\tilde{B}=0.01$ velocity trajectories follow
(at least qualitatively) their  $B=0$ counterparts, 
in the sense that they end up in the same fixed point. However, the second, third and
fourth trajectories (counting from the upper left trajectory in clockwise direction)
differ significantly from their $B=0$ counterparts. The second $\tilde{B}=0.01$ trajectory
moves apart from the $B=0$ solution simply because the latter develops a 
finite-time singularity, which is regularized by the introduction of finite 
surface tension. However, the third and fourth trajectories exhibit a quite
surprising behavior: the computed interface with $\tilde{B}=0.01$ ends up in a 
different fixed point than the exact $B=0$ solution, despite the fact that the $B=0$
solution is smooth for all time and has the asymptotic width that would be selected 
by vanishing surface tension.

\begin{figure}
\includegraphics{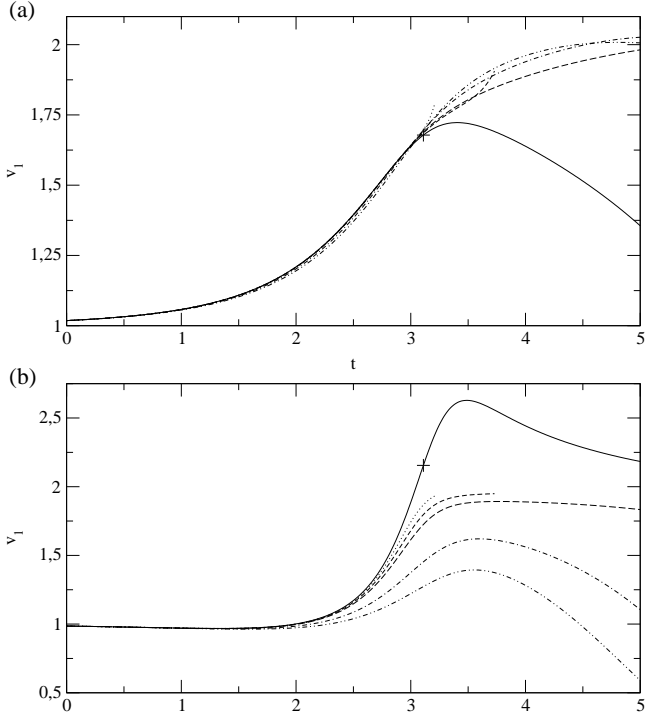}
\caption{\label{fig:tip01} Computed tip velocities for the initial condition of 
Fig.~\ref{fig:ep01} , (a)
corresponds to the central finger and (b) to the lateral finger.
The daughter singularity impact time $\tilde{t}_d$ is indicated by the $+$ symbol. The value
of $\tilde{B}$ is: 0 (solid line), 0.0002 (dotted line), 0.0005 (dashed line),
0.001 (long dashed line), 0.005 (dot-dashed line) and 0.01 (dot-dot-dashed line).
The deviations observed at late times for $\tilde{B}=0.0002$ and $\tilde{B}=0.0005$
in (b) are due to numerical noise.}
\end{figure}

In order to get further insight into this behavior we have computed the 
evolution for decreasing values of $\tilde{B}$ using the specific
 initial
pole position $\zeta^2_s(0)=20\exp(-{\rm i}\pi/6)$, with $\lambda=1/2$ and $\epsilon=0.1$.
Quadruple precision has been used when it has been necessary. 
Figure~\ref{fig:ep01}  shows its evolution for four values of the surface tension parameter,
together with the $B=0$ solution. The differences between the two interfaces
for long times are readily apparent.  When $B=0$ the finger in the central position stops growing and the
side finger wins the competition, whereas  for $B>0$ we encounter
the opposite situation---namely, the central finger surpasses the side finger and wins the
competition.   For the smaller values of $B$ the finger on the sides has not quite stopped growing when
the computation is stopped, although its tip speed shows a marked
decrease over that for $B=0$ and is 
less than that of the central
finger.  The side finger tip speed
is also decreasing at the final stage of the computation.  The tip speed trend in the limit  $B \rightarrow 0$
is further  illustrated in Fig.~(\ref{fig:tip01}). This figure  shows the tip
speed versus time of each finger  for a sequence of decreasing $B$.  The plot suggests that upon
impact of the daughter singularity the side finger velocity levels off and 
eventually decreases, whereas the velocity of the center finger is nearly unaffected
and continues to increase.  The trend is indicative of the center finger ``winning'' the competition
in the $B>0$ dynamics, while the opposite occurs for $B=0$.
Finally, it is noted that the influence of surface tension
is first felt by the smaller finger, which is the recipient
of the daughter singularity impact. Afterwards the
leading finger begins to widen, in a manner consistent with the conjecture 
in Sec.~III.  Further remarks on this point are made in Sec.~V.

\begin{figure}
\includegraphics{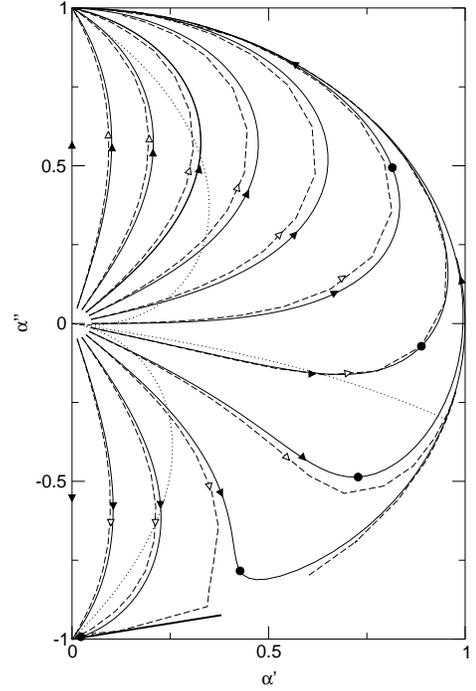}
\caption{\label{fig:proj01} Comparison between the $\tilde{B}=0$ trajectories and the projected
evolutions with $\tilde{B}=0.01$, for the initial conditions of Fig.~\ref{fig:ep01}.
The solid line corresponds to $\tilde{B}=0$ and the dashed lines to the projection
of the $\tilde{B}=0.01$ evolutions. The daughter singularity impacts are indicated
by a circle. Note that the fourth $B>0$ trajectory (as measured counterclockwise from
the bottom) reverses direction and heads toward the fixed point (-1,0).}
\end{figure}

The projection method described in the previous section has been also applied
to this case, and the results are displayed in Fig.~\ref{fig:proj01} 
in the particular case ${\tilde{B}}=0.01$.
It can be seen that for most trajectories
the projection stays close to the $B=0$ curves, even for long times. 
The daughter singularity impact still leads to $O(1)$ differences between the $B=0$
and $B>0$ solutions, although the impact does not produce changes in the outcome
of finger competition.  However,  as expected some of the trajectories (namely the third and fourth
as measured clockwise from the bottom) do indicate
significant qualitative differences in the long time evolution.
The plot provides a simple depiction of the topological inequivalence  of the $B>0$ 
and $B=0$ dynamics~\footnote{It is noted
that the $B>0$  interfacial profile and the projection profile do differ significantly
in small scale features.  This is another indication of the difficulties in using zero surface
tension solutions to describe, even qualitatively, the finite surface tension dynamics.}.  

It has been shown that the introduction of a finite $B$ has not changed the 
attractors of the problem, but it has changed their basins of attraction. Interestingly, 
in the $B=0$ case there  does not exist a single  separatrix trajectory between the 
two Saffman-Taylor attractors, but rather a finite region, corresponding to the set of 
trajectories ending in cusps, that acts as an effective separatrix. Since for
finite surface tension there are no cusps, it can be assumed that there is a
single trajectory that separates the two basins of attraction. Obviously, this trajectory
will depend on the value of the surface tension parameter. 
More precisely,  the initial condition $\zeta_s^2(0)$
corresponding to  the separatrix trajectory will be a function
of the surface tension $B$. 
To quantitatively characterize 
this set of initial conditions 
we have studied the dependence of 
the separatrix trajectory in a neighborhood of the planar interface 
fixed point as a function of $\tilde{B}$, 
using initial conditions of the form $\zeta^2_s(0)=20\exp(i\theta)$.
For a given initial condition $\zeta_s(0)$
introduce the parameter $\theta_{sep}({\tilde{B}})$, defined
as the unique value for which the evolution is attracted toward the fixed point
ST(L) when $\theta>\theta_{sep}$ and to the fixed point ST(R) when $\theta<\theta_{sep}$.
  
\begin{figure}
\includegraphics{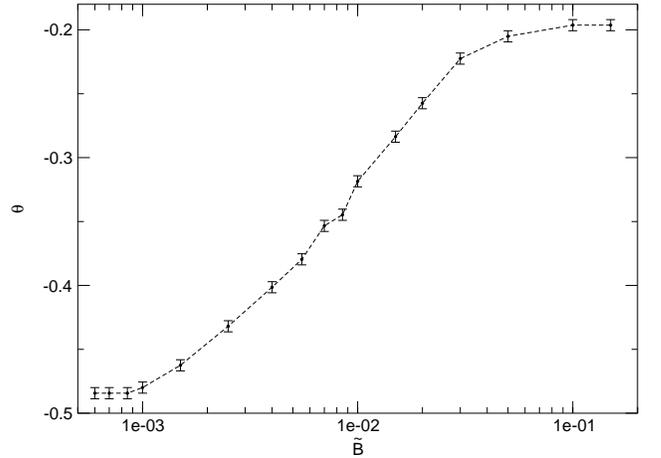}
\caption{\label{fig:sep} Plot of $\theta_{sep}$ vs. $\tilde{B}$ for  initial conditions of
the form~(\ref{fepsilon}) with $\lambda=1/2$,  $\epsilon=0.1$ and
$\zeta^2_s(0)=20\exp(i\theta)$.}
\end{figure}

Figure~\ref{fig:sep} shows the plot of $\theta_{sep}$ versus $\tilde{B}$, and it
is observed that as $\tilde{B}$ decreases, $\theta_{sep}$ saturates to a fixed value,
namely $\theta_{sep}(\tilde{B} \rightarrow 0)= -0.4843\pm0.0009$.  It is interesting to compare this value 
to the position of the separatrix region for $B=0$, which is
located between $\theta_+^{B=0}=-0.95758$ and $\theta_-^{B=0}=-1.04796$. 
The separatrix for finite $\tilde{B}$ lays outside and far away
from the separatrix region for  $B=0$, even for vanishing surface tension.
Our evidence therefore suggests that  any $B=0$ trajectory located between the
trajectories defined by $\theta_{sep}(\tilde{B} \rightarrow 0)$ and $\theta_+^{B=0}$ 
will not describe, even qualitatively, the regularized dynamics in the limit
$\tilde{B} \rightarrow 0$, since the finger that will `win' the competition 
under the $B=0$ dynamics will `lose' under the $B \rightarrow 0$ 
dynamics. Thus, there exists a positive measure
set of initial conditions of the form (\ref{fepsilon})
such that  the evolution with $B \rightarrow 0$ 
cannot be qualitatively described 
by its evolution under $B=0$ dynamics. 
This is a dramatic consequence of  the singular nature 
of surface tension on the dynamics of finger competition which is not 
related to steady state selection, but confirms the ideas of the proposed 
dynamical solvability scenario in Ref.~\cite{Paune02a}.

\section{Summary and concluding remarks} 
\label{concl}

The asymptotic theory developed in Refs.~\cite{Tanveer93,Siegel96b} predicts
the existence of regions of the complex plane where the zero surface tension
solution and the finite surface tension solution differ by $O(1)$. These
regions are the daughter singularity clusters,
and their influence is felt in the
physical interface when they are close to the unit circle. Daughter
singularities move towards the unit circle, and when their motion is not
impeded by other singularities they reach the unit circle in $O(1)$ time.
When the distance between the daughter singularity and the unit circle is
$O(B^{1/3})$ the interface can display $O(1)$ discrepancies with respect
the interface of the $B=0$ solutions. However, the asymptotic theory
does not predict the nature of the discrepancies caused by daughter singularity
impact.
 
Siegel \it et al.\rm~\cite{Siegel96b} showed numerically that 
the effect of the daughter singularity
impact on the tip (in a single-finger configuration with $\lambda < 1/2$)
was a retard in the velocity of the finger accompanied by a widening
of the finger. However, this provided small insight on the effect of
the impact in multi-finger configurations, where finger competition
could be substantially affected by the presence of finite surface tension,
as suggested in Refs.~\cite{Casademunt00,Magdaleno98,Paune02a}.

Since the precise effect of the  daughter singularity cannot be established
by the asymptotic theory it is necessary to use numerical computation
in order to establish the effects of daughter singularity on the
dynamics of the interface.
We have focused our efforts on uncovering the role of surface tension in the
dynamics of two finger configurations, which is the simplest situation
exhibiting nontrivial finger competition.
Two different types of two-finger
zero surface tension solutions have been studied. The first
type ($\epsilon=0$) does not exhibit finger competition when $B=0$ but
rather 
contains asymptotic configurations consisting
of two unequal steady fingers advancing with the same speed. These
two-finger steady state solutions form a continuum of fixed points
in the phase space of the corresponding, reduced dynamical system, 
which is structurally unstable. 
Numerical computations with small surface tension show that the
introduction of a small $B$ removes the continuum of fixed points and
triggers the competition process which was absent for $B=0$ by restoring 
the saddle-point (hyperbolic) structure of the appropriate multifinger 
fixed point. 
The second type ($\epsilon\neq0$) of two-finger solution we have
studied exhibits finger competition for $B=0$, but the numerical computation
with small $B$ has shown that the long time configuration of the
computed interface may be \it qualitatively \rm different from the
$B=0$ solution for a broad set of initial conditions,
in the sense that the finger that `wins' the competition
is not the same with and without surface tension. Thus, the presence
of surface tension  seemingly can change the outcome of finger competition even in
configurations that are well behaved and smooth for all time and whose
asymptotic width is fully compatible with the predictions of selection
theory for vanishing surface tension. This unexpected result shows
that surface tension is not only necessary to select the asymptotic 
width and to prevent cusp formation, but plays also an essential
role in multifinger dynamics through a drastic reconfiguration of the 
phase space flow structure.

Our calculations support the conjecture that impact on either the shorter or
larger finger retards the velocity of that finger, and is accompanied by the
widening of the larger finger. As a consequence, in general the outcome
of finger competition is independent of the particular finger on which the
impact first occurs, and the finger which is leading at the time of the
daughter singularity impact `wins' the competition. This recipe fails
only for interfacial configurations with very similar fingers, when not only the 
position of the finger (which finger is leading) but also the tip velocities
(a trailing finger can have for a certain time a larger velocity than the leading one)
at the impact time may play a role. 

The main conclusion of the present work is that surface tension is essential
to describe multifinger dynamics and finger competition, even when the corresponding
zero surface tension evolution is well behaved 
and compatible with selection theory. That is, we have detected singular 
effects of surface tension on the dynamics of 
finger competition that are not directly related to steady state 
selection. These can be properly interpreted in the context of an extended 
dynamical selection scenario as described in Ref.~\cite{Paune02a} where 
the reconfiguration of phase space flow by surface tension 
can be traced back to the restoring 
of hyperbolicity of multifinger fixed points.\\

\begin{acknowledgments}
We acknowledge financial suport from the Direcci\'on General de
Ense\~nanza Superior (Spain), Project No. BXX2000-0638-C02-02 (J. C.  and E. P.)
and National Science Foundation grants DMS-9704746 and DMS-0104350 (M. S.).
E. Paun\'e also ackowledges
financial support from the Departament d'Universitats, Recerca i Societat
de la Informaci\'o (Generalitat de Catalunya), and the kind hospitality of 
the Mathematics Department of the New Jersey Institute of Technology.
\end{acknowledgments}

\end{document}